\documentclass[12pt,preprint,usenatbib]{aastex}
\usepackage{amssymb}
\usepackage{amsmath}
\usepackage{graphicx}

\slugcomment{Accepted by APJ}

\shorttitle{Inflow motion in high-mass cores} \shortauthors{Wu et
al.}

\begin{document}

\title{Infall and outflow detections in a massive core JCMT 18354-0649S}

\author{Tie Liu\altaffilmark{1}, Yuefang Wu\altaffilmark{1}, Qizhou Zhang\altaffilmark{2}, Zhiyuan Ren\altaffilmark{1}, Xin Guan\altaffilmark{1,3} and Ming Zhu\altaffilmark{4}}

\altaffiltext{1}{Department of Astronomy, Peking University, 100871,
Beijing China; liutiepku@gmail.com, ywu@pku.edu.cn }
\altaffiltext{2}{Harvard-Smithsonian Center for Astronomy, 60 Garden
St., Cambridge, MA 02138, USA} \altaffiltext{3}{Current affiliation:
I. Physikal. Institut, Universit\"{a}t zu K\"{o}ln, Z\"{u}lpicher
St.77,D-50937 K\"{o}ln, Germany} \altaffiltext{4}{National
Astronomical Observatories, Chinese Academy of Sciences, Beijing,
100012}

\begin{abstract}
We present a high-resolution study of a massive dense core JCMT
18354-0649S with the Submillimeter Array. The core is mapped with
continuum emission at 1.3 mm, and molecular lines including
CH$_{3}$OH ($5_{23}$--$4_{13}$) and HCN (3--2). The dust core
detected in the compact configuration has a mass of $47~M_{\odot}$
and a diameter of $2\arcsec$ (0.06 pc), which is further resolved
into three condensations with a total mass of $42~M_{\odot}$ under
higher spatial resolution. The HCN (3--2) line exhibits asymmetric
profile consistent with infall signature. The infall rate is
estimated to be $2.0\times10^{-3}~M_{\odot}\cdot$yr$^{-1}$. The high
velocity HCN (3-2) line wings present an outflow with three lobes.
Their total mass is $12~M_{\odot}$ and total momentum is
$121~M_{\odot}\cdot$km~s$^{-1}$, respectively. Analysis shows that
the N-bearing molecules especially HCN can trace both inflow and
outflow.
\end{abstract}


\section{Introduction}

Studies of high-mass star formation have received much attention
during recent years. One of the main questions is whether massive
stars form through an accretion-disk-outflow process, similar to
low-mass counterparts (Shu, Adams, \& Lizano 1987), or via
collision-coalescence (Wolfire \& Cassinelli 1987; Bonnell, Bate, \&
Zinnecker. 1998). Studying the characteristics of massive cores at
the early stages is critical for understanding their formation
process. High-Mass Protostellar Objects (HMPOs) are precursors of
UC~H{\sc ii} regions, and represent an essential phase in high-mass
star formation (Churchwell 2002). HMPOs often have strong dust
emission and high bolometric luminosity. But their radio emission is
weak or non detectable at a level of approximately 1 mJy (Molinari
et al. 1996, 2000; Sridharan et al. 2002; Beuther, Schilke, \&
Menten. 2002; Wu et al. 2006).. Their natal clouds have not been
affected significantly by the star forming process. Thus, they
present the information about the early kinematic processes of high
mass star formation.

The dense core JCMT 18354-0649S was first detected in an ammonia
survey of high-mass star forming regions with Max-Planck-Institut
f\"{u}r Radioastronomie  (MPIfR) 100 m telescope at Effelsberg (Wu
et al. 2006), and was later confirmed by the observation with the
Submillimeter Common-User Bolometric Array (SCUBA) of James Clerk
Maxwell telescope (JCMT) (Wu et al. 2005). Another SCUBA core which
harbors a UC H~{\sc ii} region G35.4NW is located about $1\arcmin$
north of JCMT 18354-0649S. The kinetic distance of the two SCUBA
cores is 5.7 or 9.6 kpc (Wu et al. 2005), and 5.7 kpc was adopted in
this paper. Core JCMT 18354-0649S has no counterpart in radio
continuum. Multiple lines towards this core including HCN (3--2),
H$^{13}$CO$^{+}$ (3--2), and C$^{17}$O (2--1) reveal typical "blue
profile" (Wu et al. 2005), indicating that the core is undergoing
gravitational collapse (Keto, Ho \& Haschick. 1988; Zhou et al.
1993; Zhang, Ho, \& Ohashi. 1998; Wu \& Evans. 2003; Wu et al. 2005;
Fuller, Williams, \& Sridharan. 2005; Wyrowski 2006; Wu et al. 2007;
Birkmann et al. 2007; Klaassen \& Wilson 2007; Sun \& Gao 2008;
Velusamy et al. 2008). The core is also associated with a
near-infrared point source, corresponding to a star of 6-11
$M_{\odot}$ (Zhu et al.2010, submitted). Carolan et al. (2009)
observed sixteen different molecular line transitions including CO,
HCN, HCO$^{+}$ and their isotopes in this region, and modeled the
source with a chemically depleted rotating envelope collapsing onto
a central protostellar source which has evolved sufficiently to
generate a molecular outflow. All the evidence suggests that JCMT
18354-0649S is forming high mass protostellar object(s)(HMPO).
However, single-dish observations with a resolution of
$15\arcsec-40\arcsec$ can not reveal detailed kinematics in the core
at a distance as large as 5.7 kpc. In this paper we report the
results of a high resolution study with the Submillimeter Array
(SMA\footnote {Submillimeter Array is a joint project between the
Smithsonian Astrophysical Observatory and the Academia Sinica
Institute of Astronomy and Astrophysics and is funded by the
Smithsonian Institution and the Academia Sinica.}) in order to probe
the details of the core and its kinematic features. The observation
and initial results are presented in sections 2 and 3. Properties of
infall and outflow motions are discussed further in section 4, and a
brief summary is given in section 5.

\section{Observations}

The observations of JCMT 18354-0649S was carried out with the SMA in
July 2005 with seven antennas in its compact configuration and in
September 2005 with six antennas in its extended configuration. The
345 GHz receivers were tuned to 265 GHz for the lower sideband (LSB)
and 275~GHz for the upper sideband (USB).  The frequency spacing
across the spectral band is 0.8125~MHz or $\sim$1 km s$^{-1}$ for
both configurations. The phase reference center of both observations
was R.A.(J2000)~=~18$^{\rm h}$38$^{\rm m}$08.10$^{\rm s}$ and
DEC.(J2000)~=~-$6\arcdeg46\arcmin52.17\arcsec$.

In the observations with the compact configuration, Jupiter, Uranus
and QSO 3c454.3 were observed for antenna-based bandpass correction.
An amplitude offset was found on some baselines and the
baseline-based errors in bandpass were further corrected using the
point source QSO 3c454.3. QSOs 1741-038 and 1908-201 were employed
for antenna-based gain correction.
Uranus was observed for flux-density calibration.  The synthesized
beam size is $3.76\arcsec\times2.72\arcsec$ (PA=-54$\arcdeg$).

For the extended configuration, QSO 3c454.3 was used as bandpass
calibrator, QSOs 1741-038 and 1908-201 as gain calibrators, and
Uranus as a flux calibrator, respectively. The synthesized beam size
is about $1\arcsec$.

Miriad\footnote {http://carma.astro.umd.edu/miriad} was employed for
calibration and imaging. The 1.3 mm continuum data were acquired by
averaging all the line-free channels over both the 2 GHz of upper
and lower spectral bands. MIRIAD task "selfcal" was employed to
perform self-calibration on the continuum data. Since the dust
emission is weak, self-calibration with phase only was performed.
The gain solutions from the self-calibration were applied to the
line data.

The continuum data combined from both configurations yield a
synthesized beam of $1.63\arcsec\times1.28\arcsec$
(PA=-81.4$\arcdeg$), and 1 $ \sigma$ rms of 2.5 mJy in the naturally
weighted maps. HCN (3-2) and CH$_{3}$OH ($5_{23}$--$4_{13}$) were
detected in the compact configuration.

The shortest baseline in compact configuration observations is 16.5
m, corresponding to a spatial scale of $20\arcsec$. Spatial
structures more extended than this limit, such as HCN maps close to
the cloud velocity, would be filtered out. HCN (3--2) data in JCMT
archive (Wu et al. 2005; Carolan et al. 2009) were used to recover
the missing flux. The JCMT archive data were reduced using the KAPPA
and GAIA packages in the STARLINK suite. The JCMT beam size for HCN
(3--2) was $18.3\arcsec$, and the main beam efficiency was 0.69. The
combination of the SMA compact and JCMT HCN (3--2) data was done
using the task "immerge" in MIRIAD.

\section{Results}
\subsection{Dust core}
The 1.3 mm continuum images are shown in Fig.1. The left panel is
obtained in the compact array and the right panel from the combined
data of the compact and extended configurations. An elongated core
is revealed with the 1.3 mm continuum emission observed with the
compact array, and is further resolved into three condensations by
the continuum emission using the combined data from both
configurations. The three condensations are named as MM1, MM2 and
MM3. The peak position of MM1 is R.A.(J2000)=$18^{\rm h}38^{\rm
m}08.1^{\rm s}$, DEC.(J2000)=$-6\arcdeg46\arcmin52.98\arcsec$. MM2
peaks at R.A.(J2000)=$18^{\rm h}38^{\rm m}07.9^{\rm s}$,
DEC.(J2000)=$-6\arcdeg46\arcmin51.36\arcsec$, and MM3 peaks at
R.A.(J2000)=$18^{\rm h}38^{\rm m}08.05^{\rm s}$,
DEC.(J2000)=$-6\arcdeg46\arcmin51.36\arcsec$.

From the fit of an elliptical Gaussian, the core revealed by the
compact array is found to be elongated from south-east to
north-west. It has an average FWHM diameter of 0.06 pc
($\sim2\arcsec$) at a distance of 5.7 kpc, smaller than the beam
size of the compact configuration. The total integrated flux is 0.47
Jy. The total dust and gas mass can be obtained with the formula
$M=S_{\nu}D^{2}/\kappa_{\nu}B_{\nu}(T_{d})$, where $S_{\nu}$ is the
flux at 1.3 mm, D is the distance, and $B_{\nu}(T_d)$ is the Planck
function. We adopt a dust opacity
$k_{1330}$=$1.4\times10^{-2}$~cm$^{2}$g$^{-1}$ at 1.3 mm calculated
from Ossenkopf \& Henning (1994) with a dust opacity index
$\beta=2$. Here the ratio of gas to dust is taken as 100. Molecular
line CH$_{3}$OH ($5_{23}$--$4_{13}$) is detected, and its emission
peak coincides with the dust core very well (see Sec.3.2). The upper
energy level of the CH$_{3}$OH ($5_{23}$--$4_{13}$) line is 57 K
above the ground, indicating a relatively warm conditions. Assuming
$T_{d}$=57~K, a total dust and gas mass of $47~M_{\odot}$ is
derived. A beam-average gas/dust density amounts to
$2.0\times10^{6}$~cm$^{-3}$, which is larger than
$1.1\times10^{6}$~cm$^{-3}$ obtained from single-dish telescope (Wu
et al. 2005).

The three condensations (MM1, MM2 and MM3) have total integrated
flux of 0.42 Jy, leading to a total mass of $42~M_{\odot}$ (assuming
T$_{d}$ = 57~K as above). MM1 has a diameter of 0.02 pc, and a mass
of $30~M_{\odot}$. MM1 is centrally concentrated and compact, while
MM2 and MM3 are much more diffuse and extended.

With UKIRT (United Kingdom Infrared Telescope) Zhu et al. (2010, in
preparation) detected three near-infrared sources IRS1a, IRS1b, IRS1c
in H, K and L bands. Their positions are marked with crosses in
Fig.1. IRS1a ($18^{\rm h}38^{\rm m}08.135^{\rm s}$,
$-6\arcdeg46\arcmin51.57\arcsec$) lies about $1.6\arcsec$ north-east
of MM1. IRS1b ($18^{\rm h}38^{\rm m}08.026^{\rm s}$,
$-6\arcdeg46\arcmin56.24\arcsec$) and IRS1c ($18^{\rm h}38^{\rm
m}07.929^{\rm s}$, $-6\arcdeg46\arcmin55.34\arcsec$) are about
$3\arcsec$ southwest from MM1 and are much fainter.


\subsection{Gas core}
The molecular line CH$_{3}$OH ($5_{23}$--$4_{13}$) is detected in
the compact configuration. Fig.2 presents its spectrum at three
positions and the integrated emission overlaid on the 1.3 mm
continuum image. The central velocity of the CH$_{3}$OH
($5_{23}$--$4_{13}$) spectra is 96.7 km~s$^{-1}$, which is taken as
the systemic velocity of the core. The central velocity of
CH$_{3}$OH ($5_{23}$--$4_{13}$) does not shift at different
positions (see Fig.2), which should exclude rotation at the core.
The P-V diagram of CH$_{3}$OH ($5_{23}$--$4_{13}$) is shown in
Fig.3, indicating a compact gas core without rotation. The emission
center of CH$_{3}$OH ($5_{23}$--$4_{13}$) (R.A.(J2000)=$18^{\rm
h}38^{\rm m}08.092^{\rm s}$,
DEC.(J2000)=$-6\arcdeg46\arcmin52.318\arcsec$) coincides with MM1
very well. While there are no CH$_{3}$OH components corresponding
with MM2 and MM3. The deconvolved size of the gas core revealed by
CH$_{3}$OH ($5_{23}$--$4_{13}$) is $3.78\arcsec\times2.76\arcsec$
(PA=-29$\arcdeg$), comparable to the synthesized beam size of the
compact array.

The HCN (3--2) (265.886GHz) spectra obtained from the SMA compact
configuration and from the data combined from both compact and
extended configurations are presented in the left panel of Fig.4.
Both of the two spectra are averaged over a region of
$5\arcsec\times5\arcsec$, which show a redshifted absorption dip and
broad wings. The line profiles observed with the SMA and JCMT, as
well as a combination of the two are presented in the right panel of
Fig.4. All the spectra in the right panel of Fig.4 are convolved
with the JCMT beam ($18.3\arcsec$) for comparing. One can see that
the SMA compact array observations recover less than 10$\%$ of JCMT
flux around the systematic velocity, but recover more than 30$\%$
flux at the wings. The combination of the SMA and JCMT data recovers
more than 70$\%$ of the JCMT flux at all the velocity channels.

\subsection{Kinematic signatures of lines}
\subsubsection{Infall motion}
The left panel of Fig.4 shows the most prominent feature ("blue
profile") of the HCN (3--2) line at the core. The absorption gap is
more than 8 km~s$^{-1}$ wide, ranging from 93 to 101 km~s$^{-1}$.
Fig.5 presents the channel maps of the HCN (3--2) emission from 80
km~s$^{-1}$ to 109 km~s$^{-1}$ constructed from the combined data,
which is convolved with the beam of the SMA compact configuration.
The absorption is obvious in the velocity range (95,99) km~s$^{-1}$.
The absorption dip is also clearly seen in the P-V diagrams (see
Fig.6), which is much deeper than that revealed by the single-dish
observation (Wu et al. 2005).

From the left panel of Fig.4, it is clearly to see that the centeral
velocity of the absorption dip (98 km~s$^{-1}$) is redshifted from
the systematic velocity (96.7 km~s$^{-1}$) by 1.3 km~s$^{-1}$. Such
a blue asymmetric line profile where the blue emission peak is at a
higher intensity than the red one is a collapse signature of
molecular cores (Zhou et al. 1993). The spectra constructed from
JCMT data and the combined data (the right panel of Fig.4) also show
significant "blue profile", confirming the existence of infall
motions.

\subsubsection{Molecular outflow}

Besides the absorption dip, the HCN (3--2) line exhibits remarkable
broad wings extending more than 40 km~s$^{-1}$. High-velocity gas
also can be easily identified in P-V diagrams of the HCN (3--2)
emission along the direction of P.A.=15$\arcdeg$ and
P.A.=90$\arcdeg$ as shown in Fig.6. The HCN (3--2) emission obtained
from SMA compact configuration is integrated from 80 to 87
km~s$^{-1}$ for the blue lobe and from 103 to 109 km~s$^{-1}$ for
the red lobe, respectively. The contour map of the integrated flux
are shown in Fig.7. As in the channel maps (Fig.5), we can see
several clumps in each lobe in the integrated map. The integrated
HCN (3--2) emission seems to comprise an S-shaped structure from
north-east to south. Another jet-like structure extended more than
$10\arcsec$ is also seen at the west of the continuum emission
center. IRS1a seems to be the driving source of the outflow.

The southern redshifted lobe (S-lobe) comprises two clumps named
"Clump1" and "Clump2". In the north-east blueshifted lobe (NE-lobe),
two clumps are also found and named "Clump3" and "Clump4". These
clumps are distributed along the direction of the outflow and likely
to be outward gas knots. They are probably not physically related
with other stellar sources except the driving source though Clump2
is close to IRS1b and IRS1c.

\section{Discussion}
\subsection{Infall motion}
Although the HCN emission is extended over a region larger than the
compact configuration beam, the infall region is still difficult to
confine due to the contamination of the outflow. Since the size of
the gas core traced by CH$_{3}$OH ($5_{23}$--$4_{13}$) is comparable
to the compact configuration beam size, the beam size of the compact
configuration was taken as the radius ($R_{in}$) of the infall
region (Wu et al. 2009). The kinematic mass infall rate can be
calculated using dM/dt=$4{\pi}n\mu_{G}m_{H_{2}}R_{in}^{2}V_{in}$,
where $V_{in}$,  $\mu_{G}=1.36$, $m_{H_{2}}$ and
n=$2.0\times10^{6}$~cm$^{-3}$ are the infall velocity, the mean
molecular weight, the H$_{2}$ mass, and the beam-average gas/dust
density, respectively.  The infall velocity $V_{in}$ is 1.3
km~s$^{-1}$ by comparing the systemic velocity (96.7 km~s$^{-1}$)
and the velocity of the redshifted absorbing dip (98 km~s$^{-1}$) in
the HCN (3-2) spectrum (Welch et al. 1987), leading to a kinematic
mass infall rate of 2.0$\times10^{-3}~M_{\odot}\cdot$yr$^{-1}$. In
core G10.6-0.4, the redshifted NH$_{3}$ indicates large infall
velocity $5.0\pm1.7$~km~s$^{-1}$ at about 0.05 pc, and a mass infall
rate as high as 5$\times10^{-3}~M_{\odot}\cdot$yr$^{-1}$ (Keto, Ho
\& Haschick. 1987). Also with NH$_{3}$ inverse lines, Zhang \& Ho
(1997) obtained a high infall velocity $\sim3.5$~km~s$^{-1}$ within
a region smaller than 0.02 pc towards core W51e2. Large infall
velocities ($>1.5$~km~s$^{-1}$) and mass infall rates
($>1\times10^{-3}~M_{\odot}\cdot$yr$^{-1}$) were also detected
towards G10.47 and G34.26 with HCO$^{+}$ (4--3) line (Klaassen \&
Wilson 2007). In core G19.61+0.23, an infall velocity of 2.5
km~s$^{-1}$ and a mass infall rate as high as
$6.1\times10^{-3}~M_{\odot}\cdot$yr$^{-1}$ were derived (Wu et al.
2009). It seems high mass infall rate is required by high-mass star
formation. The results of the core JCMT 18354-0649S are comparable
with those of the above sources. For comparison, the $V_{in}$ from
pure free-infall assumption is also derived with the formula
$V_{in}^{2}~=~2GM/R_{in}$. The pure free-infall velocity inferred is
2.9 km~s$^{-1}$, larger than the infall velocity obtained from the
spectrum.

Wu et al. (2005) obtained a small infall velocity
($\sim0.3~km~s^{-1}$) at a radius of 4$\arcsec$. The absorption dip
of HCN (3--2) line seen by the SMA is much deeper and broader than
that observed by JCMT. However, the kinematic mass infall rate
$\dot{M}_{in}$ ($2.0\times10^{-3}~M_{\odot}\cdot$~yr$^{-1}$)
obtained here is well coincident with that obtained with JCMT (Wu et
al. 2005), $3.4\times10^{-3}~M_{\odot}\cdot$~yr$^{-1}$.

\subsection{Properties of HCN (3-2) outflow}
The column density of HCN at each velocity channel in each outflow
lobe can be obtained through (Garden et al. 1991):
\begin{equation}
N_{HCN}(v) =
\frac{3k}{8\pi^{3}B\mu^{2}}\frac{exp[hBJ(J+1)/kT_{ex}]}{(J+1)}\frac{(T_{ex}+hB/3k)}{1-exp(-h\nu/kT_{ex})}\int\tau_{v}dv
\end{equation}
Where $v$ is the central velocity of the channel relative to the
systemic velocity, the rotational constant B=44.315976 GHz and
permanent dipole moment $\mu=3$ debye for HCN, the velocity channel
width is smoothed to be $1~km~s^{-1}$. Assuming HCN emission in the
line wings to be optically thin and excitation temperature of
T$_{ex}$=30 K (Wu et al. 2004), the optical depth $\tau_{v}$ can be
derived with the equation:
\begin{equation}
\tau_{v} =
\frac{kT_{r}(v)}{h\nu}(\frac{1}{exp(h\nu/kT_{ex})-1}-\frac{1}{exp(h\nu/kT_{bg})-1})^{-1}
\end{equation}
where $T_{r}(v)$ is the excess brightness temperature of HCN(3--2)
emission at $v$. Adopting $X_{HCN}=[HCN]/[H_{2}]=1\times10^{-10}$
(Carolan et al. 2009), the mass of each lobe at $v$ can be
calculated with:
\begin{equation}
M(v) = X_{HCN}^{-1}\mu_{G}m_{H_{2}}D^2{\int}N_{HCN}(v)d{\Omega}
\end{equation}
where D, $\Omega$ are the cloud distance and the solid angle. Thus
the total mass of each lobe is given by $M=\sum$$M(v)$, the total
momentum by $P=\sum$$M(v)v$, and the energy by
$E={\frac{1}{2}}\sum$$M(v)v^2$. The dynamical timescale $t_{dyn}$ is
estimated as $R/V_{max}$, where R is the outflow extent,
and $V_{max}$ is the maximum velocity of the outflow lobe. The
mechanical luminosity L, and the mass-loss rate $\dot{M}$ are
calculated as L=E/t, $\dot{M}=P/(tV_{w}$), where the wind velocity
$V_{w}$ is assumed to be 500 km~s$^{-1}$ (Lamers et al. 1995).
The derived parameters are listed in Table.1. The total mass,
momentum, energy of the three lobes are 12~$M_{\odot}$, 121
$M_{\odot}\cdot$ km s$^{-1}$ and $1.3\times10^{46}$~erg,
respectively. The average dynamical timescale is about
$1.6\times10^{4}$ yr, and the total mass-loss rate
$1.6\times10^{-5}~M_{\odot}\cdot$~yr$^{-1}$.
The outflow is massive with parameters similar to that of IRAS
05274+3345E and the other outflows detected towards five massive
star formation regions (Zhang et al. 2007b; Klaassen \& Wilson
2008).

The Position-Velocity diagram at the left panel of Fig.6 shows that
at low velocities, the NE-lobe and S-lobe both have compact
morphology near the core center. At higher velocities the southern
lobe becomes further away from the center. As shown in Fig.7, the
outflow axis traced by Clump1 and Clump2 differs from that traced by
Clump3 and Clump4. The different outflow orientations in the large
and small scales may be attributed to the precession of the outflow
axis (Su et al. 2007). From the right panel of the P-V diagram, a
high-velocity component (V~$<$ 85 km~s$^{-1}$) with velocities
decreasing with distance from the protostar, and a second component
tracing the low-velocity material (V~$>$ 85 km~s$^{-1}$) extending
about $15\arcsec$ along the axis of the W-lobe are clearly seen.
Such convex spur PV structure was also revealed in a simulation of a
pulsed jet driven outflow (Lee et al. 2001).


\subsection{HCN --- tracer of both inflow and outflow motions}
HCN is among the most abundant molecular species with a high
critical density larger than $10^{6}$~cm$^{-3}$ (for HCN (1--0))
(Carolan et al. 2009), and is believed to trace dense molecular
cores. HCN is detected in both low mass class 0 and I sources (Park,
Kim, \& Ming. 1999; Yun et al. 1999), and high-mass hot cores
(Boonman et al. 2001).

HCN is thought as a good tracer of inflow motions (Wu \& Evans.
2003). The infall asymmetry in the HCN spectra is found to be more
prevalent, and more prominent than in any other previously used
infall tracers such as CS (2--1), DCO$^{+}$ (2--1), and
N$_{2}$H$^{+}$(1--0) during a survey toward 85 starless cores (Sohn
et al. 2007). Among the small group of pre- and protostellar objects
in L1251B, infall signature was also detected in the HCN emission
(Lee et al. 2007). HCN also traces inflow motions very well in
massive star-formation regions. Wu and Evans. (2003) found 12
sources showing "blue profile" in the HCN lines during a
spectroscopic survey of 28 massive cores with water maser. Besides
HCN, other nitrogen bearing molecules such as N$_{2}$H$^{+}$ are
also tracers of inflow motions (Tsamis et al. 2008; Schnee et al.
2007; Crapsi et al. 2005). Recently inverse P Cygni profile of CN
line in hot cores was found (Zapata et al. 2008; Wu et al. 2009).
These results suggest nitrogen bearing molecular species be good
tracers of inflowing motions in star-formation regions.

Outflows traced by HCN are often detected not only in low-mass
star-formation regions but also in massive star-formation regions
(Bachiller, Guti\'{e}rrez, \& P\'{e}rez. 1997; Choi 2001; Su et al.
2007; Zhang et al. 2007a). HCN outflow of the core JCMT 18454-0649S
is another good sample. Additionally, Zhu et al. (2010 in
preparation) and Cyganowski (2008) found excess emission at 4.5
$\micron$ at the position of source IRS1a, which is close to the
center of the NE-lobe and S-lobe. Such excess emission at the 4.5
$\micron$ band could be shock-excited. In IRAS 20126+4104, HCN
emission is also found to be closely related to the shock-excited
near-IR H$_{2}$ knots and was identified to be associated with shock
wings (Su et al. 2007). The inner clumps (Clump1 in the S-lobe and
Clump3 in the NE-lobe) of the core JCMT 18354-0649S should also be
coincident with shocks. In fact, models have already demonstrated a
dramatic increase of HCN molecules, during the intense interaction
between outflow and ambient gas, or slow shock front (Mitchell 1984;
Nejad, Williams, \& Charnley. 1990). In this process sulfur and
nitrogen react with hydrocarbons to produce various compounds,
wherein HCN abundance gets higher than the rest of the products
(Nejad, Williams, \& Charnley. 1990). Thus HCN may trace the outflow
even better than sulfur containing molecules.

\section{Summary}
Both dust continuum at 1.3 mm and CH$_{3}$OH emission detected with
SMA reveal a compact core in JCMT 18354-0649S. The core observed
with the compact configuration has a mass of $47~M_{\odot}$ and an
average density of $2.0\times10^{6}$~cm$^{-3}$. With the combination
of the compact and extended configurations, the core is resolved to
three condensations with a total mass of $42~M_{\odot}$.

HCN (3-2) spectra exhibit an infall signature in this region. The
red shifted absorption seen in the SMA observation is deeper and
broader than that in the JCMT observation. The infall rate is
$2.0\times10^{-3}~M_{\odot}\cdot$~yr$^{-1}$. High velocity gas is
detected in HCN (3-2) emission. The outflow has three lobes and
their total mass is 12~$M_{\odot}$ and momentum of 121
$M_{\odot}\cdot$ km s$^{-1}$. The average dynamical timescale and
the total mass-loss rate are about $1.6\times10^{4}$ yr and
$1.6\times10^{-5}~M_{\odot}\cdot$~yr$^{-1}$, respectively. All the
findings indicate a high-mass protostar is forming via rapid
accretion. Our results suggest that nitrogen bearing molecules
especially HCN are good for probing both infall and outflows.

\section*{Acknowledgment}
\begin{acknowledgements}
We are grateful to the SMA staff. We also thank Dr. Shengli Qin. for
his help with data reduction and discussion. This work was funded by
Grants of NSFC No 10733030 and 10873019.
\end{acknowledgements}

\begin{table}[c]
{
\begin{center}
\caption{Outflow parameters of each lobe.\label{tbl}}
\begin{tabular}{cccccccccr}
\hline \hline
outflow             & V$_{max}$   &  $t_{dyn}$     &  Mass        &  momentum       &    Energy     &  $L$    &  $\dot{M}_{out}$   \\
\quad               & km~s$^{-1}$ & (10$^4~yr$)    &  ($M_{\odot}$)   & ($M_{\odot}$ km~s$^{-1}$)  & (10$^{45}~erg$) & ($L_{\odot}$)  & ($10^{-6}M_{\odot}$~yr$^{-1}$)  \\
\hline
southern lobe       &  11.7      &  1.4              & 3.3  & 29  & 3.0  &  1.8  &  4.0        \\
north-eastern lobe  &  15.3      &  1.0              & 4.2  & 42  & 4.5  &  3.6  &  8.0      \\
western lobe        &  15.3      &  2.3              & 4.6  & 50  & 5.5  &  2.1  &  4.0      \\

\hline
\end{tabular}
\end{center}
}
\end{table}

\clearpage
\begin{figure}
\includegraphics[angle=-90,scale=.5]{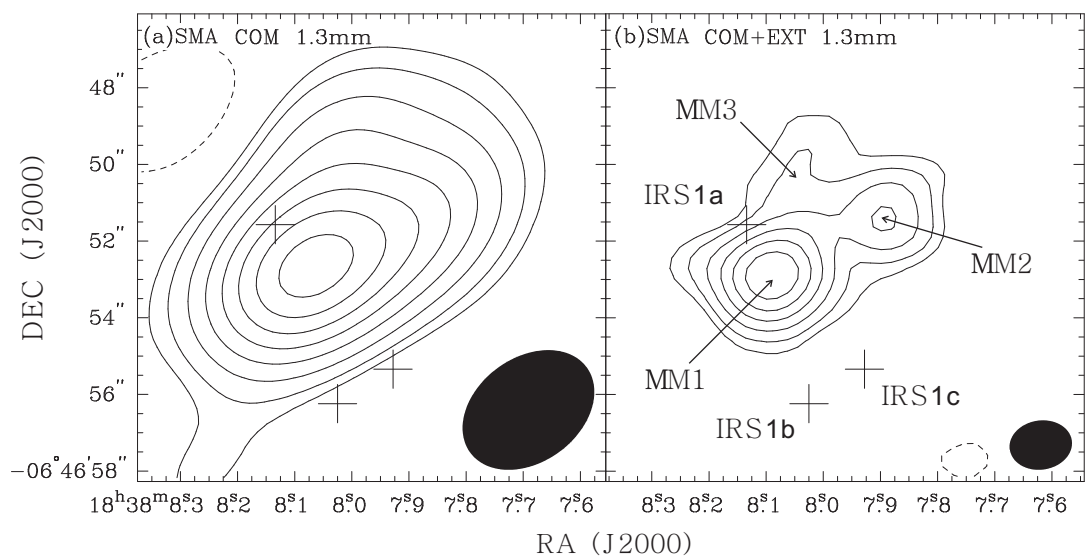}
\caption{The 1.3 mm continuum emission towards JCMT 18354-0649S. The
left one is obtained with the compact configuration. The rms level
is 3 mJy~beam$^{-1}$ (1 $\sigma$). The contours are at -6, 3, 6, 12,
21, 33, 48, 66, 87$\sigma$. The right panel gives the contours of
the continuum emission combined from both configurations. The rms
level is 2.5 mJy~beam$^{-1}$ (1 $\sigma$) and the contours are at
-6, 3, 6, 12, 21, 33, 48$\sigma$. The three near-infrared sources
are marked with crosses.}
\end{figure}

\begin{figure}
\includegraphics[angle=90,scale=.60]{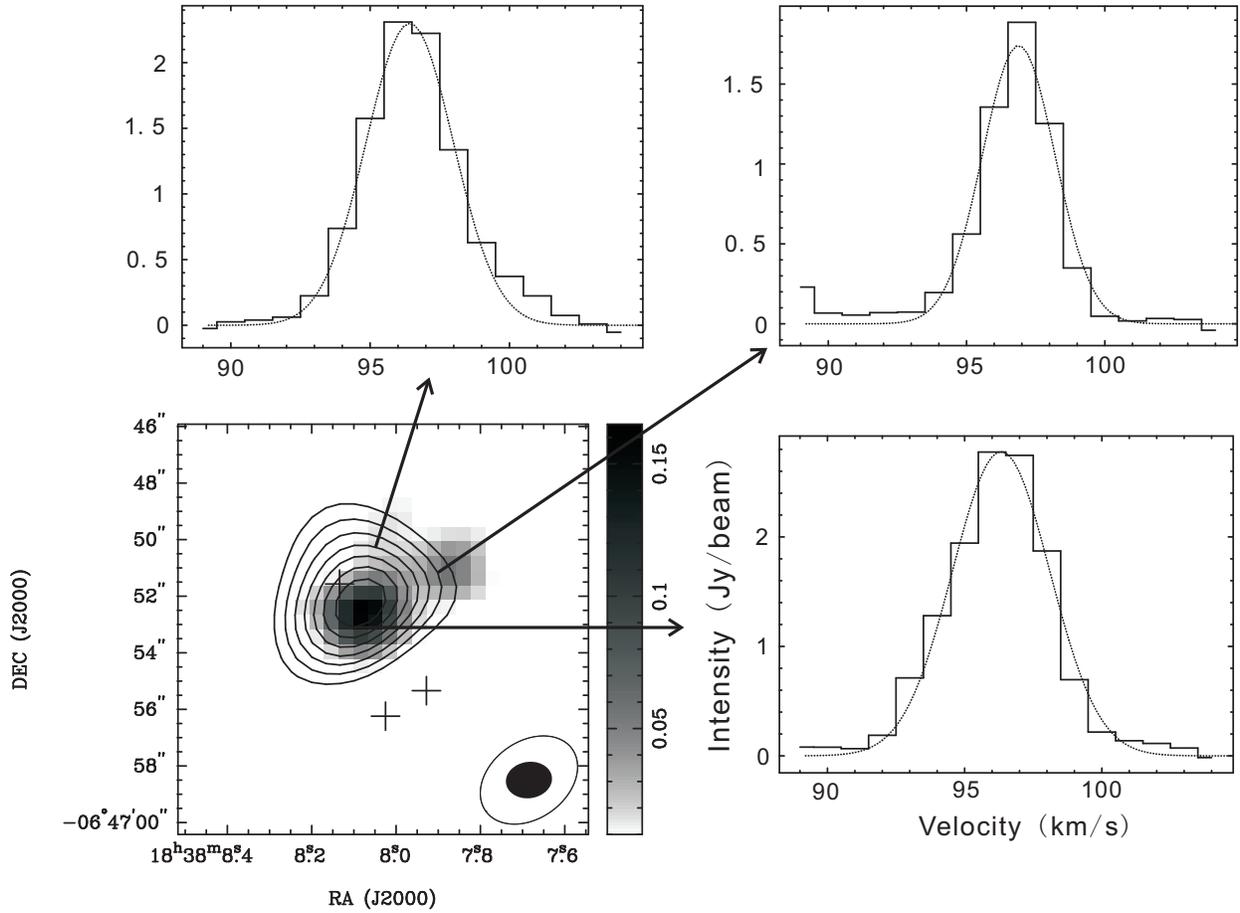}
\caption{The lower-left panel is the contours of the CH$_{3}$OH
integrated intensity overlaid on the 1.3 mm continuum image (grey
scale). The contours start from 30$\%$ in steps of 10$\%$ of the
peak emission (15 Jy~beam$^{-1}\cdot$km~s$^{-1}$). The three
near-infrared sources are marked with crosses. The beam-averaged
spectrum of CH$_{3}$OH at three positions are presented in the other
panels. The gaussian fit towards each spectrum is shown with solid
lines.}
\end{figure}

\begin{figure}
\includegraphics[angle=-90,scale=.40]{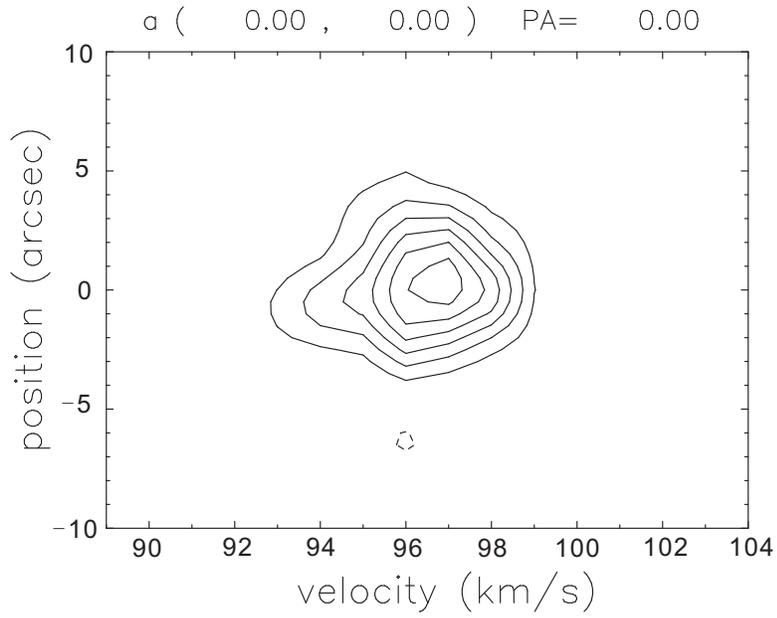}
\caption{Position-velocity diagrams of CH$_{3}$OH along a P.A. of
0$\arcdeg$. The contour levels are from 15\% to 90\% in steps of
15\% of the peak intensity in both panels. The intensity at the peak
is 3.73 Jy~beam$^{-1}$.}
\end{figure}

\clearpage

\begin{figure}
\begin{minipage}[c]{0.5\textwidth}
  \centering
  \includegraphics[width=75mm,height=60mm,angle=0]{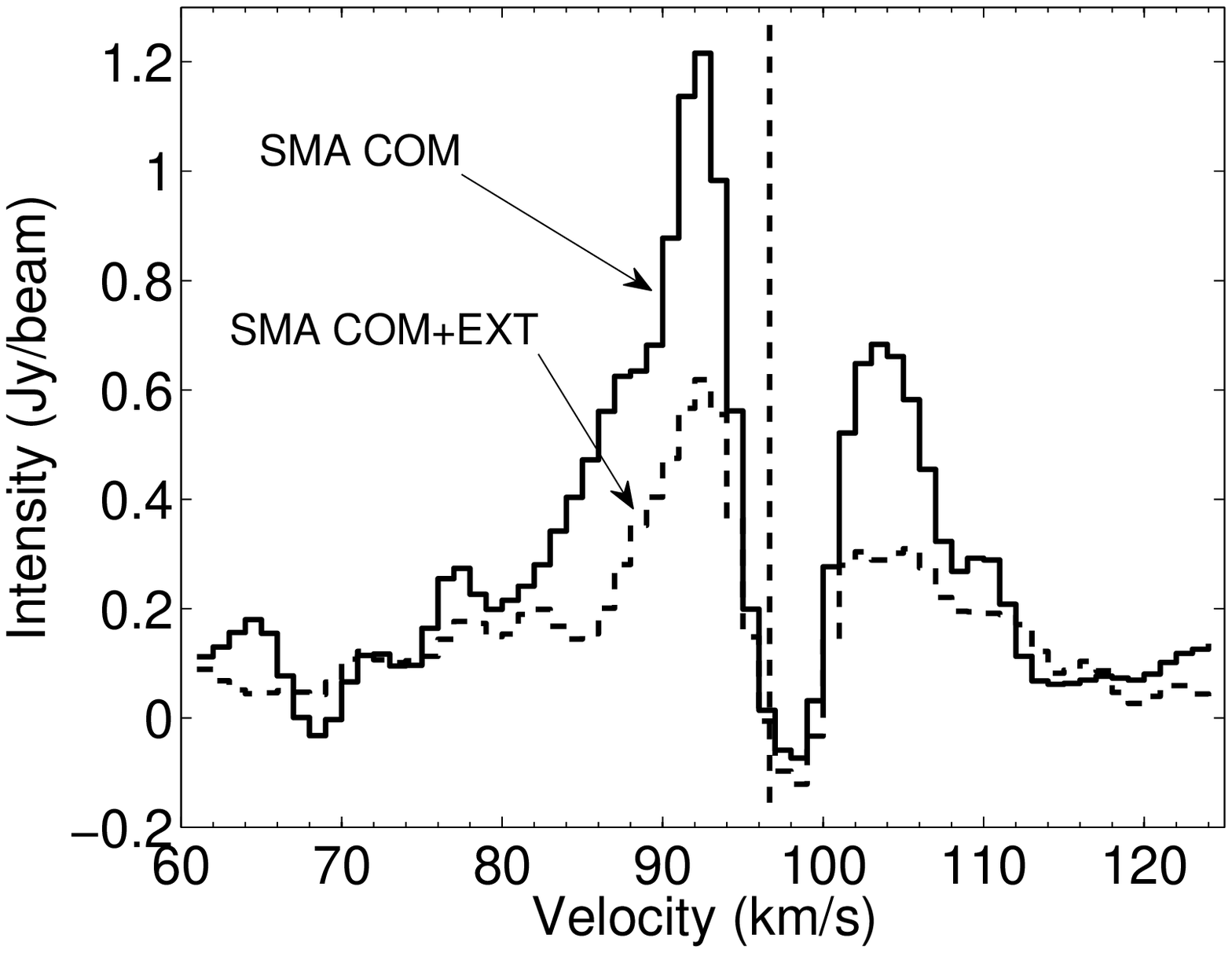}
\end{minipage}
\begin{minipage}[c]{0.33\textwidth}
  \centering
  \includegraphics[width=75mm,height=60mm,angle=0]{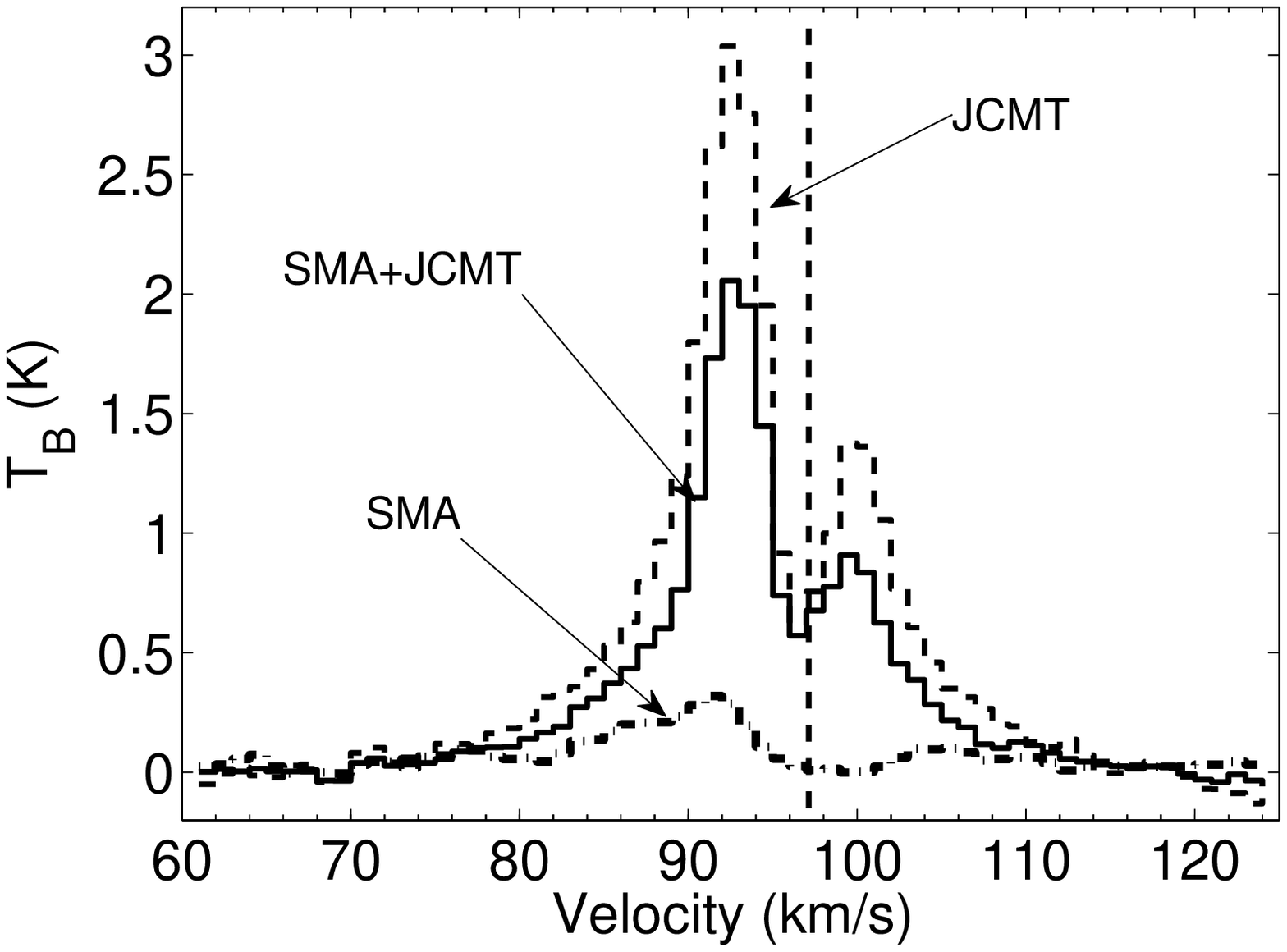}
\end{minipage}
\caption{The HCN (3-2) spectra. Left: the solid black line exhibits
the spectrum constructed from SMA compact array and the dashed line
shows the spectrum obtained from combining the compact and extended
data together. Both of the two spectra are integrated over a region
of $5\arcsec\times5\arcsec$. Right: the solid black line shows the
spectrum constructed from the combined SMA and JCMT data, which is
convolved with the JCMT beam ($18.3\arcsec$); the dashed line shows
the spectrum from JCMT only; the dash-dotted gray line shows the
spectrum obtained with the SMA compact array and convolved with the
JCMT beam. The vertical dashed lines in both panels mark the
position of the systematic velocity (96.7 km~s$^{-1}$).}
\end{figure}

\begin{figure}
\includegraphics[angle=-90,scale=.60]{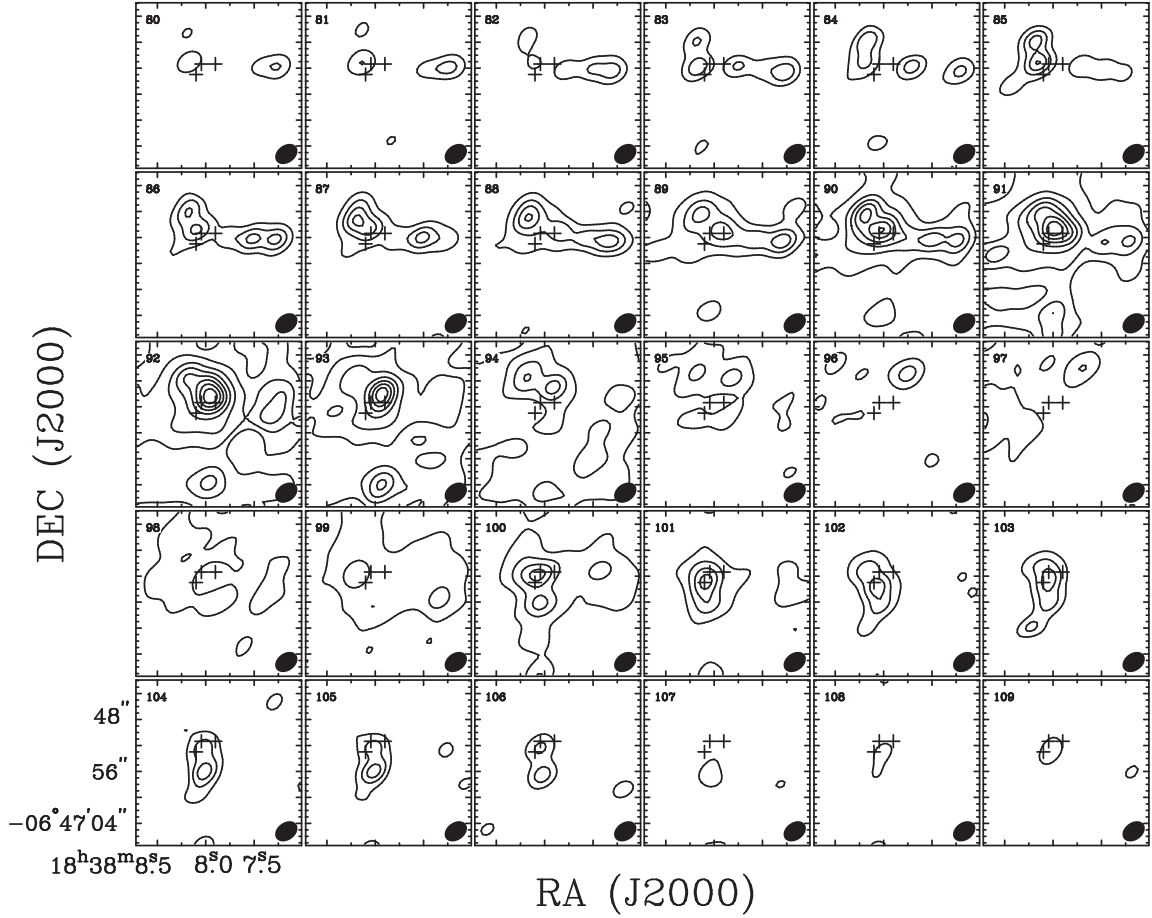}
\caption{Combined JCMT and SMA HCN (3-2) channel maps from 80
km~s$^{-1}$ to 109 km~s$^{-1}$, which is convolved with the beam of
the SMA compact configuration ($3.76\arcsec\times2.72\arcsec$,
PA=-54$\arcdeg$). The contours are in steps of 0.5 Jy~beam$^{-1}$
(3$~\sigma$) from 0.5 Jy~beam$^{-1}$ (3$~\sigma$). The velocity of
each channel is plotted at the upper-left of each panel, and the
beam size at the lower-right. The positions of MM1, MM2 and MM3 are
marked with crosses.}
\end{figure}

\begin{figure}
\begin{minipage}[c]{0.5\textwidth}
  \centering
  \includegraphics[width=60mm,height=70mm,angle=270]{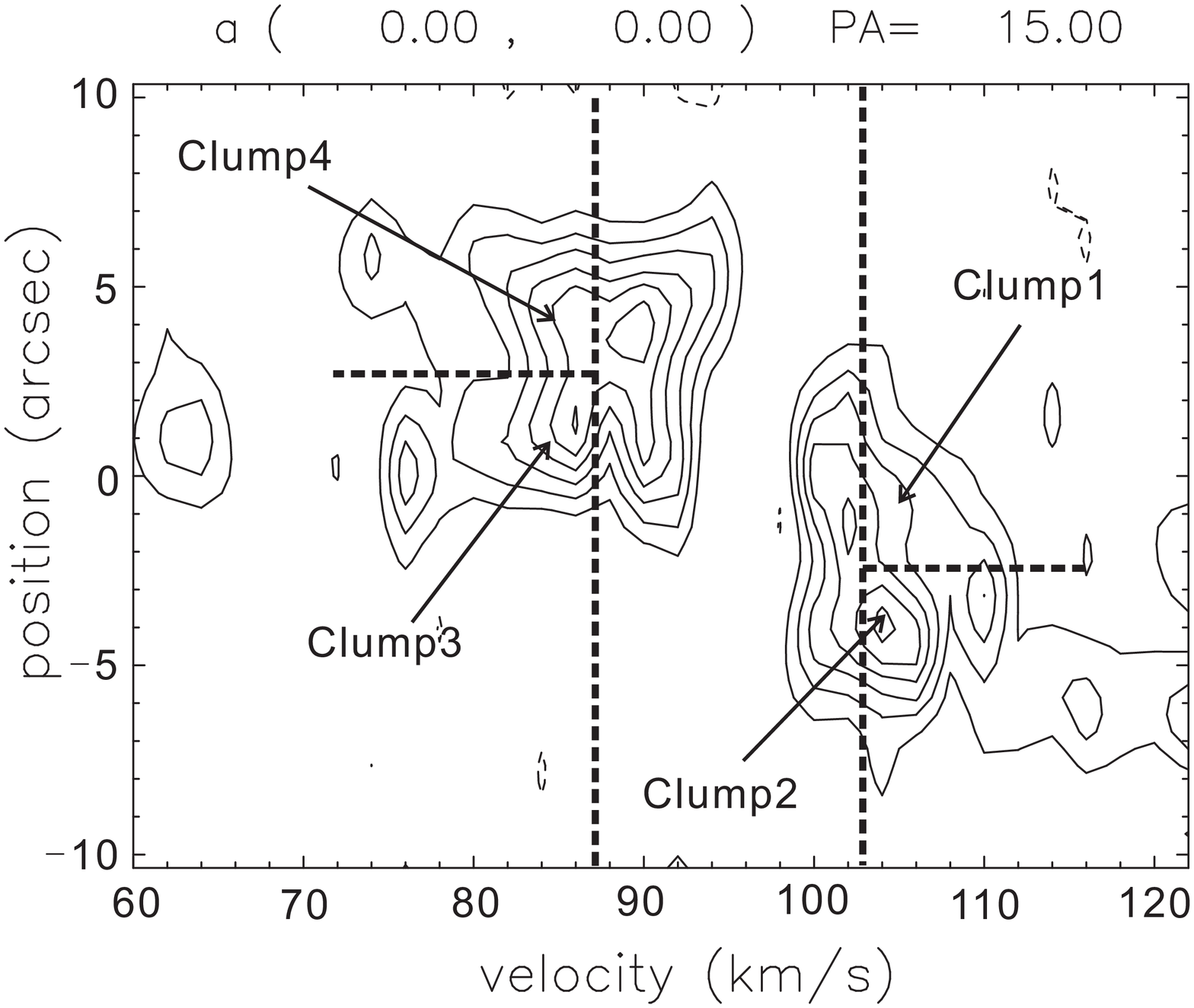}
\end{minipage}
\begin{minipage}[c]{0.33\textwidth}
  \centering
  \includegraphics[width=60mm,height=70mm,angle=90]{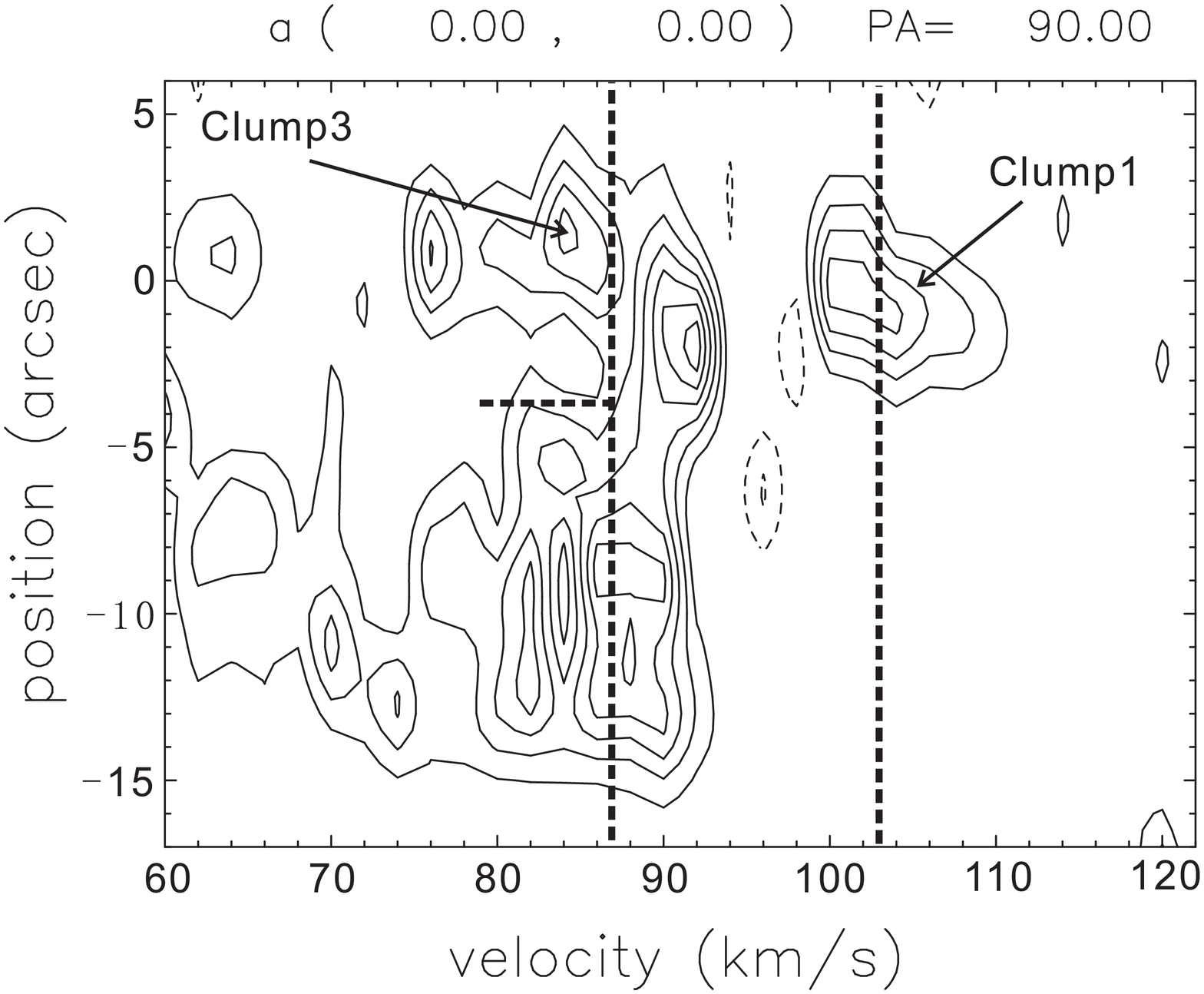}
\end{minipage}
\caption{Position-velocity diagrams of HCN (3-2) outflow observed by
the SMA along a P.A. of 15$\arcdeg$ (left panel) and 90$\arcdeg$
(right panel). The image is smoothed to 2 km~s$^{-1}$ velocity
resolution. The contour levels are from 15\% to 90\% in steps of
15\% of the peak intensity in both panels. The peak is 1.84
Jy~beam$^{-1}$ in the left panel and 1.92 Jy~beam$^{-1}$ in the
right panel. The four clumps are labeled by solid lines with arrows.
The clumps are distinguish by the thick dashed lines.}
\end{figure}

\begin{figure}
\includegraphics[angle=90,scale=.50]{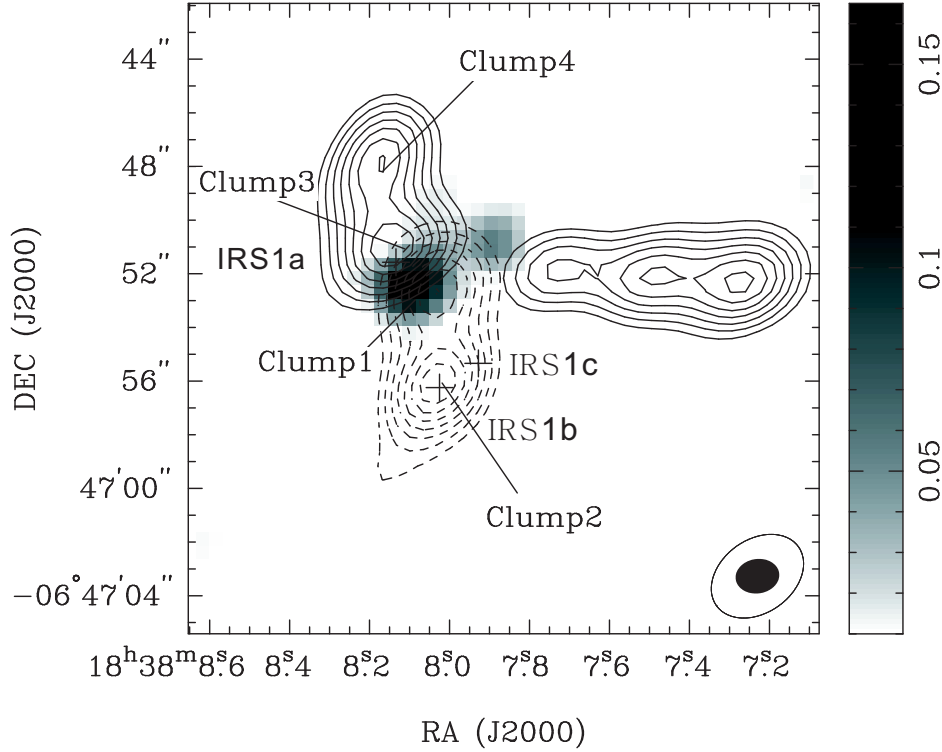}
\caption{The high-velocity HCN (3-2) intensity contours overlaid on
the 1.3 mm continuum image, integrated from 80 to 87 km~s$^{-1}$ for
the blueshifted lobes (solid contours) and from 103 to 109
km~s$^{-1}$ for the redshifted lobe (dashed contours), with contours
from 30$\%$ in steps of 10$\%$ of the peak emission. The peak is
8.55 Jy~beam$^{-1}$$\cdot$km~s$^{-1}$ for the blueshifted lobes and
7.16 Jy~beam$^{-1}$$\cdot$km~s$^{-1}$ for the redshifted lobe. The
empty and solid ellipses in the lower-right corner represent the
synthesized beams of HCN (3-2) emission and 1.3 mm continuum
emission combined from both configurations, respectively.}
\end{figure}

\clearpage

\end{document}